\documentstyle[aasms4,12pt,epsf]{article}
\begin{document}

\title{ Near Infrared Observations of a Redshift 4.92 Galaxy:
Evidence for Significant Dust Absorption \altaffilmark{1}}

\altaffiltext{1}{Based on observations obtained at the W.M. Keck
Observatory} 

\author{B.T. Soifer\altaffilmark{2}, G. Neugebauer\altaffilmark{2}, M.
Franx\altaffilmark{3}, K. Matthews\altaffilmark{2}, G.D.
Illingworth\altaffilmark{4}}

\altaffiltext{2}{Palomar Observatory, 320-47, Caltech, Pasadena, CA 91125}

\altaffiltext{3}{Kapteyn Institute, P.O. Box 800, NL-9700 AV, Groningen,
The Netherlands}

\altaffiltext{4}{University of California Observatories / Lick
Observatory, Board of Studies in Astonomy and Astrophysics, University of
California, Santa Cruz, CA 95064}

\received{1997 December 12}
%\revised{1998 March 20}
\accepted{1998 May 15}

\begin{abstract}

\gdef\Msun{M_{\odot}}

Near-infrared imaging and spectroscopy have been obtained  of the
gravitationally lensed galaxy at $z=4.92$ discovered in HST images by
Franx et al. (1997). Images at 1.2, 1.6 and 2.2$\mu$m show the same arc
morphology as the HST images.  The spectrum with resolution $\lambda /
\Delta\lambda\sim 70$ shows no emission lines with equivalent width
stronger than 100\AA\ in the rest frame wavelength range 0.34$\mu$m to
0.40$\mu$m. In particular, [OII]3727\AA\ and [NeIII]3869\AA\ are not seen.
The energy distribution is quite blue, as expected for a young stellar
population with the observed Ly $\alpha$ flux.  The spectral energy
distribution can be fit satisfactorily for such a young stellar population
when absorption by dust is included. The models imply a reddening 0.1 mag
$< E(B-V) <$0.4 mag.  The  stellar mass of the lensed galaxy lies in the
range of 2 to 16 $\times 10^9$ $\Msun$. This is significantly higher than
estimates based on the $HST$ data alone. Our data imply that absorption by
dust is important to redshifts of $\sim 5$.

\medskip

\end{abstract}

\keywords{ Galaxies: formation; starburst}

\section{Introduction}
\label{sec:introduction}

The study of galaxies at high redshifts has been revolutionized with the
introduction of the Hubble Space Telescope and the new generation of large
ground-based telescopes.  Detecting and studying such systems at high
redshift is essential to understanding how normal galaxies form and
evolve.  Franx et al. (1997) have recently reported the serendipitous
discovery of a field galaxy at a redshift $z=4.92$ that has been
gravitationally lensed into several arc components, including a highly
magnified fold arc,  by  an intervening cluster of galaxies  (CL1358+62)
at $z=0.33$.  Franx et al. present HST imaging and Keck spectroscopy of
this highly-magnified lensed system.  Because of the redshift, the HST
images at $R$ ($F606W$) and $I$ ($F814W$) correspond to wavelengths of
1023\AA\ and 1375\AA\ in the rest frame of the background objects, and
thereby sample only the far UV portion of the light emitted from these
galaxies.  In order to study these objects at wavelengths where nearby
galaxies have been studied, we observed the brightest (fold) arc of the
lensed galaxy at near infrared wavelengths using the W.M. Keck Telescope.
In the discussion we adopt the same cosmological parameters as in Franx,
et al., i.e. $H_0 =$ 50 Km/s/Mpc and $q_0=$0.5.

\smallskip

\section{Observations and Data Reduction}
\label{sec:observations}

The observations reported  here were made in April 1997 in photometric
conditions using the near infrared camera (NIRC) at the f/25 forward
Cassegrain focus of the W.M.  Keck Telescope.  The instrument is described
in detail by Matthews and Soifer (1994).  It has a 256$\times$256 InSb
array with 0.15$\arcsec \times$0.15$\arcsec$ pixels for a 38$\arcsec
\times$ 38$\arcsec$ field of view.

\smallskip

Sets of images centered on the brightest portion of the arc (locations B
to C, described by Franx et al. 1997) were obtained at 1.25$\mu$m ($J$),
1.65$\mu$m ($H$) and 2.2$\mu$m ($K$).  The broadband $J$, $H$ and $K$
images were made with total exposures of 1620 s per filter.  In addition a
set of images was taken in a 1.2\% bandpass filter centered at a
wavelength of 2.297 $\mu$m (rest wavelength 3881\AA ), which contained the
redshifted [NeIII] 3869\AA\ line).  The total exposure time at this
wavelength was 2700 s.  The imaging data were obtained in sets of 60 s
exposures, where the center of the field was moved in a 3$\times$3 grid
pattern 10$''$ on a side.  The centers of successive grids were moved by
1--2$''$ between each set.   Bright cluster galaxies in the frames were
used to accurately register the successive frames for coaddition.

\smallskip

Sky and normalized flat-field frames were created from the data at each
wavelength separately, taking the clipped mean of nine frames centered in
time on the frame of interest. The seeing, as determined from several
stellar images in the field, was approximately 0.5$\arcsec$ --
0.6$\arcsec$ FWHM.  The data were calibrated using calibration stars of
Persson (1997).

\smallskip

In addition to the images, low resolution grism spectra were obtained of
selected sources within the arc.  These spectra were obtained over the
wavelength range 1.5 -- 2.4$\mu$m (0.25 -- 0.40 $\mu$m in the rest
frame).  The slit was 0.7$\arcsec$ wide yielding a spectral resolution of
$\lambda$/$\Delta\lambda \sim 70$.  Slit position angles of $80^\circ$ and
$145^\circ$ were used.  The brightest  portion of the fold arc C, the knot
and the northwest end (Franx et al.  1997) was contained in the slit for
the PA $80^\circ$ spectrum, while the slit was aligned with the arc for
the PA $145^\circ$ spectrum.  For each position angle the integration time
was 3000 s.  The knot at the NW end of arc C was chosen because it was the
brightest of the lensed knots, although the data of Franx et al show that
the strength of Ly$\alpha$ is clearly lower at that position (the
equivalent width of Ly$\alpha$ at this location drops to 1/3 of its value
elsewhere in the arc).

\smallskip

\section{Results}
\label{sec:results}

%\subsection { Images}

The $J$ image of the field of CL1358+62 is shown in Figure 1a.  Locations
B and C in the nomenclature of Franx et al. (1997) are labeled.  The arc
is seen in the $J$, $H$ and $K$ images over the entire spatial range in
which it is visible in the HST images.   The arc was also detected in the
narrowband image at a level consistent with the $K$ band continuum image
and shows that the [NeIII] line does not contribute significantly to the
flux (equivalent width $< $50 \AA ) in the arc.

In order to directly compare the Keck images with the HST image we have
convolved the HST $I$ image with the measured seeing of the Keck data.  In
Table 1 we report photometry at $I$,  $J$, $H$ and $K$ in three beams;
position 1 is a circular beam including the brightest peak in the arc at
position C,  position 2 includes the full arc from B to C with the beam
shaped to mimic the arc but excluding the brightest knot, and position 3
is the sum of the first two beams. Figure 1b shows the shape of the beam
of position 3.

To derive  photometry that is less sensitive to the presence of the bright
elliptical galaxy north-northeast of the fold arc, we first determined the
surface brightness profile of this galaxy in each of the  images using an
ellipse fitting program.  We subtracted the resulting model distribution
{}from the images.  This produced a flat background at the location of the
arc that reduced the uncertainties in the photometry.  The $J$ band image
formed in this way is shown in Figure 1b.  To minimize the uncertainties
in the photometry, we made the sky beam as large as possible while
avoiding the field objects near the arc.  The photometry obtained with and
without the elliptical galaxy removed from the image was indistinguishable
at all wavelengths. Because we felt the results were more robust, the
photometry  with the elliptical galaxy removed is reported in Table 1.

%\subsection { Spectra}

The spectrum at PA $80^\circ$, taken with the bright knot of the fold arc
C centered in the slit, shows that the continuum was detected at low
signal-to-noise ratio.  No convincing detection of the arc was made in the
spectrum at PA $145^\circ$. No lines were detected in either spectrum over
the wavelength range 0.25 -- 0.40 $\mu$m in the rest frame of the object.
This corresponds to an upper limit on the equivalent width of 100 \AA\ for
any emission line over the wavelength range 0.34 -- 0.40 $\mu$m and a
limit of 200 \AA\ in the equivalent width of any emission line over the
wavelength range 0.25--0.34$\mu$m (the higher limit at shorter wavelengths
is a result of the strong OH airglow emission in the H band).

Since there is at best only marginal evidence for any color difference
between positions 1 and 2, we adopt the highest signal-to-noise
measurements of the sum of these apertures (position 3) to determine the
intrinsic energy distribution of the source galaxy.  This was determined
by measuring the colors $I - [{\lambda}]$ at position 3 and subtracting
these colors from the intrinsic $I_{AB,814}= 24.0$ mag of the lensed
galaxy G1 determined by Franx et al. (1997).  This leads to intrinsic
(unmagnified) magnitudes for G1 of 22.4, 22.3, and 21.5 mag at $J$, $H$
and $K$ respectively.  We do not believe that the quality of the data
warrant interpreting the results in terms of color variations across the
lensed object. In the discussion that follows the derived quantities refer
to the intrinsic (unlensed) galaxy.

\smallskip

\section{Stellar Population Models}
\label{sec:models}

The near infrared wavelengths correspond to rest frame UV and blue in the
lensed galaxy, with the $J$, $H$, and $K$ images being at 2150, 2790, and
3720 \AA , respectively.  The broadband colors potentially suffer from
contamination by strong emission lines.  In the $K$ band the strongest
potential contamination is due to [OII]3727\AA\ emission.  The limit of
100\AA\ on the rest frame equivalent width of the [OII] line in the
spectrum of the bright knot of C limits the contamination of the $K$
continuum to $<$20\% of the total observed flux in the band.  This limit
is consistent with that seen in young star--forming galaxies at $z\sim 1$
by Cowie et al.  (1995), who find an average rest--frame equivalent width
in [OII] of 60\AA. The limit on [OII] equivalent width limits the increase
in the spectral index due to line contamination in $K$ to $\Delta\beta<$
0.3, where the spectral index is defined as  $F_\lambda \propto
\lambda^{\beta}$. In the $H$ band, the strongest potential contaminant is
MgII 2800\AA, while in the J band it is CIII] 1909\AA. Our limit on the
MgII line is not particularly significant, while we do not have spectra
covering the CIII] wavelength.  The CIII] and MgII lines are not seen in
the nuclear spectra of nearby starbursts such as NGC 4214 (Leitherer et
al. 1996) or blue compact galaxies such as Haro 2 (Thuan 1990), while our
limit on [OII]3727\AA\ is consistent with the strongest lines seen in
nearby blue galaxies by Gallagher et al.  (1989).  Thus we conclude that
the effect of line emission on the derived colors  is at most comparable
to the uncertainties in the observations.

The slope of the continuum is well characterized by $F_\lambda \propto
\lambda^{-1.63}$. Although this is fairly blue, it not as blue as models
predict for young, ionizing starburst populations (e.g., Leitherer and
Heckman 1995). Most of their models predict a slope $< -2.2$ for ages
smaller than $10^7$ years (as needed for the observed Ly $\alpha$ -- see
below).  The redder colors we observe can be produced either by dust, or
by the presence of older stars. In both cases the mass of the stellar
population would be seriously underestimated if the masses are estimated
{}from the bluest passband. To address this issue, we made fits to the
spectral energy distribution using the models by Bruzual and Charlot
(1993, 1996).  We considered models for populations with instantaneous
starbursts and with constant star formation rates and have included the
effect of reddening.

The results are shown in Figure 2. Instantaneous starburst models with
ages between $3\times 10^5$ and $10^7$ years are shown in Figures 2a and
2c. The age of the burst population has to be lower than $10^7$ years to
provide the ionizing flux for the Ly $\alpha$ emission.  The IMF was
assumed to be a Salpeter law.  Figure 2a shows models with no reddening
fit to the  data including the I photometry, while figure 2c shows models
fit to the data where reddening is a free parameter.  The models can
provide good fits with a dust reddening $E(B-V)$  between 0.26 and 0.37
mag, but do not fit the data satisfactorily when reddening is not
included.  The masses of the stellar population were in the range of 5 to
$16\times 10^{9}$ $\Msun$.

Models with continuous star formation can also provide a good fit to the
data once reddening is considered. Figures 2b and 2d show models with ages
up to $10^9$ years.  We do not consider models with ages $> 10^9$ years
since this is approximately the Hubble time at z$=$4.92.  As with the
instantaneous burst models, the models in figure 2b assume no reddening,
while in figure 2d reddening is a free parameter in the fit, and these
models provide signficantly better fits than the dust free models. The
required reddening $E(B-V)$ lies between  0.15 and 0.39 mag. The required
star formation rates lie between 130 and $4\times 10^4$ $\Msun yr^{-1}$.
The  total stellar masses range from 7 to $30\times 10^{9}$ $\Msun$, for
ages from $10^7$ to $10^8$ years, respectively.  The mass becomes
progressively larger with even greater ages.  As can be seen in Figure 2d,
the models with ages $\ge  10^8$ years predict a strong Balmer break, and
this significantly reduces the quality of the fit.  At an age of 10$^9$
yr, the total chi square for the dust free model is 40, while that
containing dust is 4.6. The observed equivalent width of Ly $\alpha$
between 7 and 21 \AA\ does not provide a strong constraint on these
continuous star forming models.

The new data result in much higher mass estimates than the earlier
estimates (Franx et al. 1997) which were derived from the rest-frame
far-UV flux (observed $HST$ $I_{814}$).  The masses are increased because
the fluxes in the new redder passbands are significantly higher than
predicted by the (dust free) models used by Franx et al.  An additional
dependency with these new mass estimates is on the form of the extinction
curve used.  The Calzetti et al. (1994) extinction curve is rather grey.
If we use an SMC extinction curve,  we find that the absorption and
stellar  masses  are decreased by a factor of about two.

We can also estimate masses by ignoring the extinction and simply using
the reddest flux, i.e., the observed $K$ band flux, or restframe $U$.  We
obtain masses of 1.1 to $5\times 10^9$ $\Msun$ for the instantaneous burst
models,  and masses between 1.6 and  $9\times 10^9$ for the continuous
star formation models. These estimates are roughly one quarter of the
estimates derived by using the Calzetti et al. (1994) extinction curve,
and about half the estimate derived from the SMC extinction curve. We
conclude therefore that the uncertainty in the mass estimate is about a
factor of four.

A realistic assessment of the likely mass is given by the geometric mean
of the no-reddening values and the values  based on the Calzetti et al.
(1994) extinction curve:  masses between $2\times 10^9$ and $9\times
10^{9}$ $\Msun$ for instantaneous bursts, and between $3\times 10^9$ and
$16\times 10^9$ $\Msun$ for continuous star formation models.

\section{Discussion}
\label{sec:discussion}

The discussion of the last section has shown that the lensed galaxy has
significant extinction, and we have derived new larger estimates for the
stellar mass.  The high stellar masses imply that the galaxy has built up
a very dense center. The knot, which has an effective radius of 130 pc,
contains roughly half the mass, i.e., 1.2 to $8\times 10^{9}$ $\Msun$
(Franx, et al, 1997).  Assuming an isothermal profile, we derive a
velocity dispersion of the knot between 100 and 260 km s$^{-1}$.  Our new
results have therefore strengthened the case that this young galaxy at $z
= 4.92$ has already managed  to build up a very dense core.

The resulting dynamical timescale for the knot is a few times $10^6$ year,
comparable to the age derived from a starburst model.  The dynamical
timescale for the galaxy as a whole is at least 10 times higher. This
exceeds the lifetime of ionizing O stars. Combined with the fact that the
emission and absorption line properties observed by Franx et al. (1997)
indicate the presence of a strong wind which locally depletes the
interstellar medium in this galaxy, it is likely that we are seeing young,
short lived condensations of star formation.  Over time, these knots
would  brighten  progressively across the galaxy, as suggested earlier by
Lowenthal et al (1997).

The infrared photometry has demonstrated that dust is very likely present
in the galaxy.  This conclusion clearly depends on the accuracy of the
stellar evolutionary models. The reddening is estimated to be between
$0.1$ mag$< E(B-V) < 0.4$ mag, depending on the population model and
extinction curve.  These values are comparable to those found in nearby,
low metallicity, starbursts (e.g., Meurer et al 1997. This implies that
the galaxy has been able to produce a significant amount of metals, but
the observations to date are inadequate to derive metallicities.

We note that similar results were derived by Ellingson et al. (1996) for
MS1512-cB58 and for galaxies in the Hubble Deep Field by Sawicki and Yee
(1998). The optical and infrared fluxes of MS1512-cB58 at $z = 2.72$ imply
a young age ($10^7$ years) and a significant extinction ($E(B-V) \sim 0.3$
mag). The constraints on the age of the young population were stronger
because the observations extended beyond the Balmer break.  In the case of
the  Hubble Deep Field, Sawicki and Yee have found that  spectral energy
distributions with reddening of $E(B-V)\sim 0.3$mag provides better fits
to the observations of $z > 2$ Lyman Break galaxies than do dust free
models. The best studied high redshift galaxies have significant amounts
of dust while evidence for dust exists for other high redshift galaxies
based on sparser color information (e.g., Meurer et al. 1997, Pettini et
al. 1997). If the lensed galaxies are typical, then a reddening between
$E(B-V)\approx 0.2 - 0.3$ mag should be expected for most high redshift
galaxies.

\acknowledgments 

We thank  W. Harrison for assistance with the observations, and S.E.
Persson for providing photometric standards in advance of publication.
The W.M. Keck Observatory is operated as a scientific partnership between
the California Institute of Technology, the University of California and
the National Aeronautics and Space Administration. It was made possible by
the generous financial support of the W.M. Keck Foundation.  Infrared
astronomy at Caltech is supported by grants from the NSF and NASA.
Support from STScI grant GO05989.01-94A is also gratefully acknowledged.
This research has made use of the NASA/IPAC Extragalactic Database which
is operated by the Jet Propulsion Laboratory, Caltech, under contract with
NASA.

% Tables next, one per page.
%%%%%%%%%%%%%%%%%%%%%%%%%%%%%%%%%%%%%%%%%%%%%%%%%%%%%%%%%%%%%%%%%
\clearpage
\begin{table}

%\centerline {Table 1}
%\caption {Table 1}

%\centerline 
\caption{Photometry of Locations in lensed arc in CL1358+62}

\smallskip
\begin{tabular}{c c c c c c c }   
\tableline\tableline

Position & Beam & I\tablenotemark{a} & J & H & K \\
&  & mag & mag & mag & mag  \\
\tableline

1 & circ \tablenotemark{b} &   22.88$\pm0.04$  & 21.7 $\pm$0.2 & 
21.6$\pm$0.3 & 21.3 $\pm$0.3 \\

2 & arc \tablenotemark{c}& 22.79$\pm0.04$ & 21.84$\pm$0.20 &
21.40$\pm$0.30 & 20.54 $\pm$0.18 \\

3 & irr \tablenotemark{d}&     22.11  $\pm$0.03&  20.92  $\pm$0.11& 20.83
$\pm$0.22 & 19.99$\pm$0.14 \\

%\tableline

\tablenotetext{a} {The magnitude referred to here as I is that from the
HST image in the F814W filter from the data of Franx, et al. 1997.}

\tablenotetext{b} {Photometry refers to a 1.8$''$ diameter circular beam,
centered on object C (in nomenclature of Franx et al.)}

\tablenotetext{c} {Photometry refers to arc shaped beam that includes
object B and the arc between B and C (in nomenclature of Franx et al.)  }

\tablenotetext{d} {Photometry  refers to an irregular shaped beam that
includes both objects B and C and the arc between them (the sum  of
positions 1 and 2)}

\end{tabular}

\end{table}

\clearpage

%\begin{table}

%\centerline {Table 1}
%\caption {Table 1}

%\centerline 
%\caption{ Colors in CL1358+62 lens}

%\smallskip
%\begin{tabular}{c c c c c c c }   
%\tableline\tableline

%Position\tablenotemark{a}  & I-J & J-H &  J-K \\
%&  mag & mag &  mag  \\
%\tableline

%1 & 1.18$\pm$0.2 &  0.1$\pm$0.3   &  0.4$\pm$0.3  \\

%2 &  0.95$\pm$0.2 & 0.4$\pm$0.4 &  1.3$\pm$0.3\\ 

%3 &  1.19$\pm$0.1  & 0.1$\pm$0.2 &  0.9 $\pm$0.2\\

%\tableline
%\tablenotetext{a} {Position refers to same locations and beams as in Table 1}
%\tablenotetext{b} {Spectral index is defined as F$_{\lambda} \sim %\lambda^{\beta}$ where $\beta$ is defined between the rest wavelengths %corresponding to the photometry at the appropriate infrared wavelengths}

%end{table}
%\clearpage

\clearpage

\begin{figure}
\caption{
Image of Lens Arc associated with the cluster CL1358+62 in the J band
obtained with the W.M.Keck Telescope.  The top panel is the image of the
field with the orientation indicated in the image, the bottom panel is the
same image with the lensing galaxy removed. Locations B and C from Franx,
et al. are indicated in the top image, while the outline of the
photometric aperture C is shown in the bottom image.}

\label{fig images of CL1358+62}
\end{figure}

\begin{figure}
\caption{
The energy distribution for the lensed system in CL1358+62, plotted as
luminosity density vs. rest wavelength compared with instantaneous and
continuous star formation models from Burzual and Charlot (1993, 1996).
The left panels, represent instantaneous models of ages  10, 30 and 100
$\times 10^5$ years, the top panel being unreddened, the bottom panel
having a reddening E(B-V) between 0.23 and 0.33 mag.  The right panels
represent continuous star formation models with ages of 3, 10, 30, 100,
300, 1000 and $10,000\times 10^5$ years. The top panel is unreddened
models, the bottom panel having reddening between 0.15 and 0.39 mag to
best fit the data, with the older models requiring less reddening. For
both sets of models, ages below the youngest listed age are
indistinguishable.}

\label{fig:energy distribution/models} \end{figure}

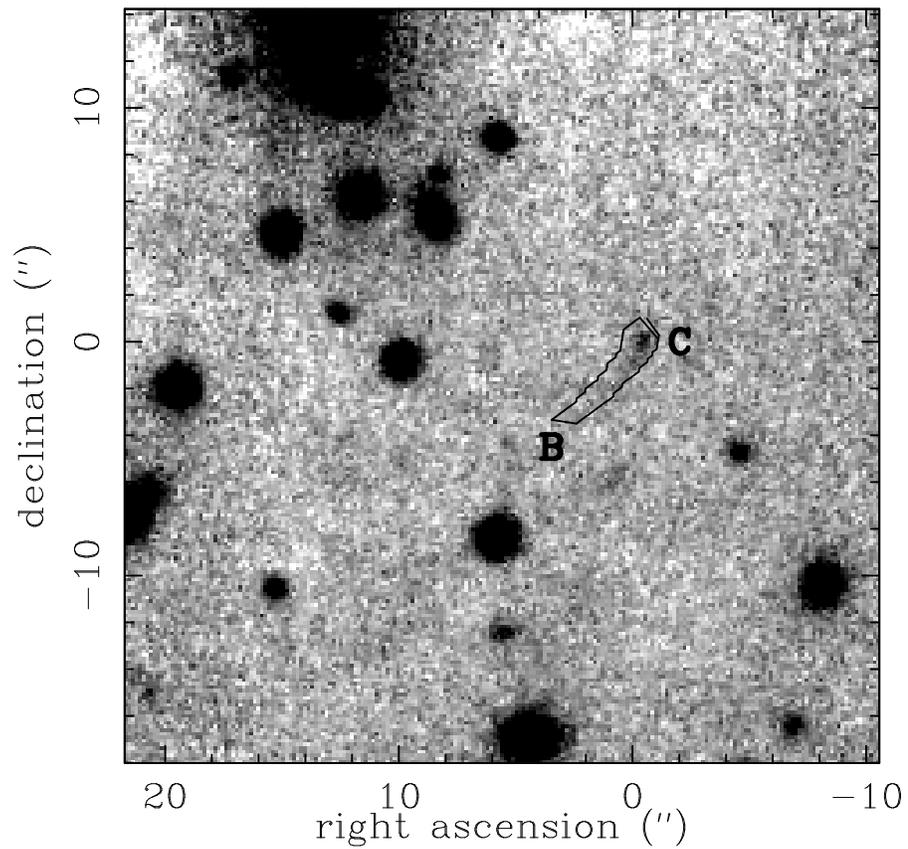
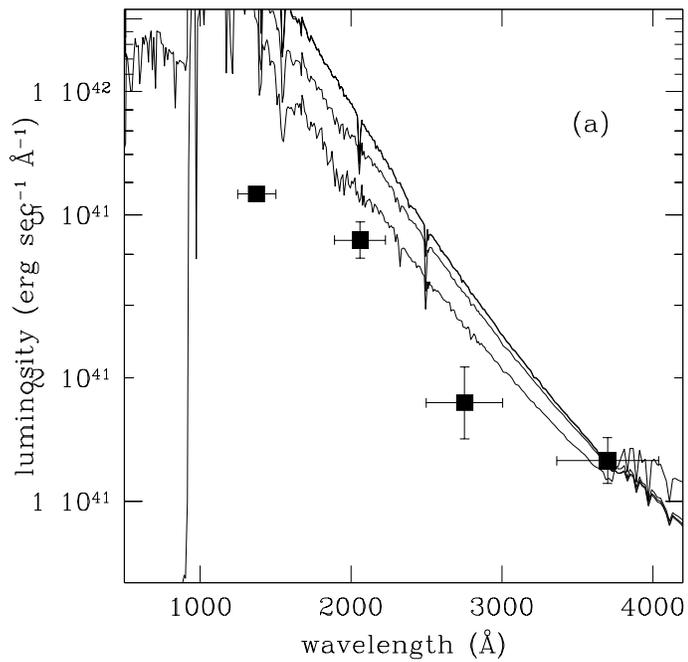
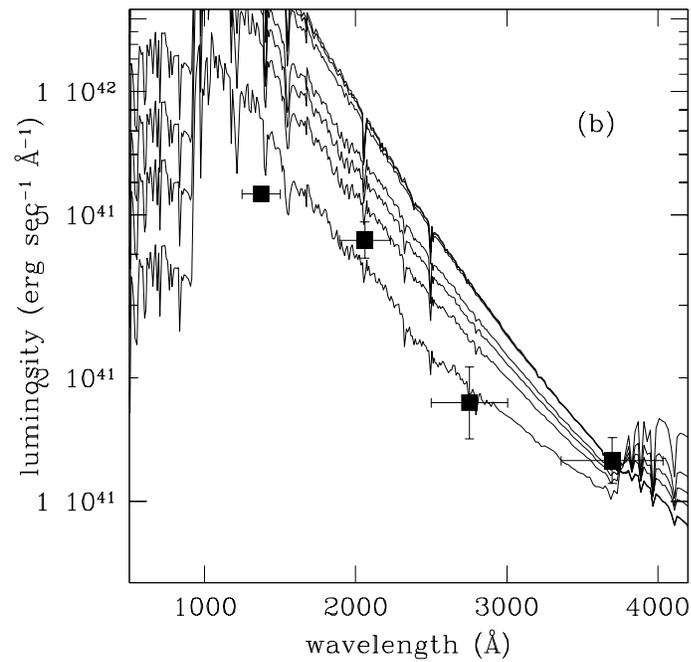
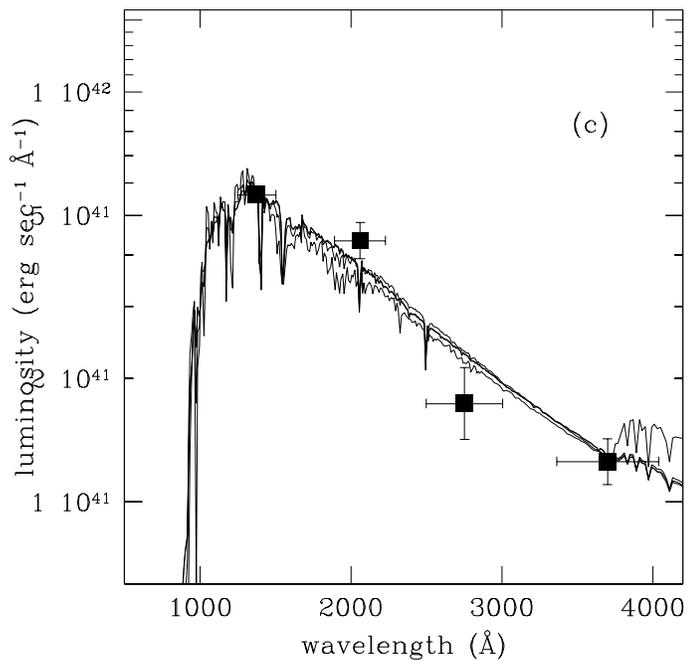
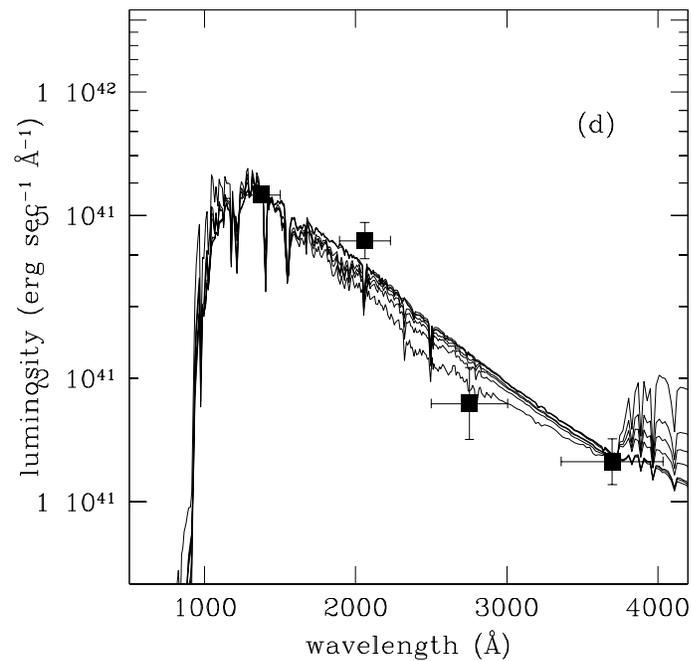
\end{document}